\documentclass[aps,prl,showpacs,twocolumn]{revtex4-1}

\usepackage{graphicx}
\usepackage{longtable}
\usepackage[ansinew]{inputenc}

\begin{document}

\title{Two-Site Kondo Effect in Atomic Chains}

\author{N. Néel}
\author{R. Berndt}
\affiliation{Institut für Experimentelle und Angewandte Physik,
Christian-Albrechts-Universität zu Kiel, D-24098 Kiel, Germany}
\author{J. Kröger}
\affiliation{Institut für Physik, Technische Universität Ilmenau, D-98693
Ilmenau, Germany}
\author{T. O. Wehling}
\author{A. I. Lichtenstein}
\affiliation{Institut für Theoretische Physik, Universität Hamburg, 
D-20355 Hamburg, Germany}
\author{M. I. Katsnelson}
\affiliation{Institute for Molecules and Materials, Radboud University 
Nijmegen, NL-6525 AJ Nijmegen, The Netherlands}

\begin{abstract}
Linear CoCu$_n$Co clusters on Cu(111) are fabricated by means of atomic 
manipulation. They represent a two-site Kondo system with tunable interaction.
Scanning tunneling spectroscopy reveals oscillations of the Kondo temperature 
$T_{\text{K}}$ with the number $n$ of Cu atoms for $n\geq 3$.
Density functional calculations show that the Ruderman-Kittel-Kasuya-Yosida 
interaction mediated by the Cu chains causes the oscillations. Calculations 
find ferromagnetic and antiferromagnetic interaction for $n=1$ and $2$, 
respectively. Both interactions lead to a decrease of $T_{\text{K}}$ as
experimentally observed.
\end{abstract}

\pacs{68.37.Ef,72.15.Qm,73.20.Fz}

\maketitle

Magnetic atoms with partially filled $d$ or $f$ shells induce strong electron 
correlations, which cause spectacular effects such as Mott metal-insulator 
phase transitions, heavy-fermion behavior or the occurrence of high 
temperature superconductivity. The rich physics of these systems is due to the 
interplay of local and nonlocal correlation effects. Local correlations are 
due to Coulomb interaction which makes the probability of an electron to hop 
into an unoccupied $d$ or $f$ orbital at a site $i$ depend on the number of 
electrons already present at that site. Nonlocal correlations arise from 
electrons propagating from one site $i$ to another site $j$ with Coulomb 
scattering at both sites. In this way, the Ruderman-Kittel-Kasuya-Yosida (RKKY) interaction \cite{mru_54,*tka_56,*kyo_57} can cause magnetic correlations between distant atoms and its interplay with (local) Kondo physics is believed  to be responsible for magnetic instabilities in heavy-fermion systems.

Two interacting magnetic atoms are the smallest possible solid state structure 
to display nonlocal magnetic phenomena due to these correlations. They are 
consequently an important model system and may also be viewed as a keystone in 
a bottom-up approach to magnetic nanotechnology. In single-impurity systems, 
correlations between delocalized band electrons and more localized electrons 
at the impurity define a characteristic low-energy scale referred to as Kondo 
temperature $T_{\text{K}}$, and manifest themselves as an Abrikosov-Suhl 
resonance of width $\text{k}_{\text{B}}T_{\text{K}}$ near the Fermi level.
Interactions between two impurities may alter this energy scale and in the 
limit of strong interactions lead to new regimes including ferromagnetically 
locked impurity spins or an interimpurity singlet 
\cite{cja_81,bjo_87,*bjo_88}.

In this work the characteristic low-energy scale is explored as a function of the interaction between two impurities.
A Co atom on a Cu(111) surface is connected to a second Co atom by a chain of Cu atoms.
By evaluating the widths of the Abrikosov-Suhl resonances from scanning tunneling spectra we find that $T_{\text{K}}$ is significantly lower for short CoCu$_n$Co ($n=1,2$) than for longer chains ($3\leq n\leq 6$).
Starting from $n=3$, $T_{\text{K}}$ exhibits even-odd oscillations with the number of Cu atoms.
Combining density functional 
theory (DFT) and perturbative renormalization group (RG) calculations we show 
that the observed behavior is due to magnetic interactions between the Co 
adatoms. 
The observed oscillations of  $T_{\text{K}}$ are shown to result from the RKKY interaction mediated by the Cu chains. Overall, both ferromagnetic and antiferromagnetic exchange interactions between the impurities are shown to  reduce $T_{\text{K}}$ compared to the  single-impurity Kondo scale $\tilde T_{\text{K}}$.

A few scanning tunneling microscope (STM) experiments previously addressed 
magnetic interactions between individual adsorbed atoms (adatoms). The magnetic 
hysteresis of two interacting Co atoms on a Pt(111) surface was investigated 
as a function of the interatomic distance \cite{lzh_10}. Kondo physics was not 
addressed in that work. The antiferromagnetic coupling between a Kondo 
Co atom and an unscreened Fe atom on Cu$_2$N surface induced a splitting of the 
Kondo resonance similar to Kondo splitting of single atom in external magnetic 
field \cite{ott_09}. One Co adatom separated by a few Cu layers from an entire magnetic Co film can be similarly understood in terms of single-impurity Kondo physics \cite{tuc_08}. In both cases, none of the quantum phase transitions associated with the two-impurity Kondo problem \cite{rbu_08} can arise. A disappearance of the Kondo effect or a reduction of 
$T_{\text{K}}$ in small Co$_2$ and Ni$_2$ clusters have been reported 
\cite{wch_99,vma_02}. Moreover, the Kondo effect of two Ni atoms on Au(111) 
was found to be fully developed as soon as the Ni atoms were not in 
nearest-neighbor positions \cite{vma_02}. Finally, a broadening of the 
Abrikosov-Suhl resonance has been extracted from tunneling spectra of Co atoms 
in next-nearest neighbor positions and used to determine their exchange 
interaction $E_{\text{ex}}$ \cite{pwa_07}. However, the method used in 
Ref.\,\cite{pwa_07} to relate $T_{\text{K}}$ and $E_{\text{ex}}$ has been 
questioned \cite{emi_09}. While in these cases the interaction between 
magnetic impurities was short-ranged and mediated by the substrate or by 
direct contact of the magnetic atoms, an intermediate chain of nonmagnetic 
atoms may affect the interaction \cite{obr_08}. 
The present work focuses on the impact of such a chain on the Kondo effect.
For the first time the long-range interaction of two impurities is explored over the entire range from weakly coupled Kondo impurities to the regime where the interimpurity exchange coupling and $\tilde T_{\text{K}}$ are of the same order of magnitude.
\begin{figure}
\includegraphics[width=55mm]{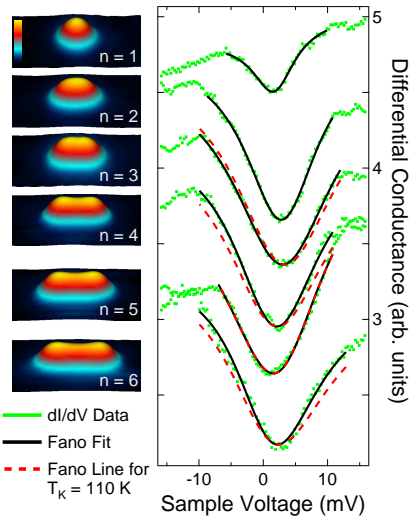}
\caption{(Color online) 
Left: Pseudo-three-dimensional representations of constant-current STM images
($38\,\text{\AA}\times 21\,\text{\AA}$) of linear CoCu$_{n}$Co clusters 
($n=1 \ldots 6$). Right: $\text{d}I/\text{d}V$ spectra recorded above the Co 
atoms of the chains. Green dots are experimental data, solid lines indicate 
Fano line shapes fit to the data. The dashed lines are Fano lines for 
$T_{\text{K}}=110\,\text{K}$, which corresponds to a Co adatom attached to a
Cu$_4$ chain.}
  \label{fig1}
\end{figure}
	
The experiments were performed with a STM operated at $7\,\text{K}$ and $10^{-9}\,\text{Pa}$. Cu(111) surfaces and chemically 
etched W tips were prepared by Ar$^+$ bombardment and annealing. Single Co 
atoms were deposited at $\approx 10\,\text{K}$ using an electron beam 
evaporator while single Cu atoms were transferred from the tip to the sample 
\cite{lli_05}. The adatoms were chemically identified by the presence (Co) or 
absence (Cu) of the Abrikosov-Suhl resonance. Spectroscopy of the differential 
conductance ($\text{d}I/\text{d}V$) was performed by a lock-in technique using 
a modulation amplitude of $1\,\text{mV}_{\text{rms}}$. Only tips which reproduced the known spectrum of a single Co adatom were used.

\begin{figure}
  \includegraphics[width=70mm]{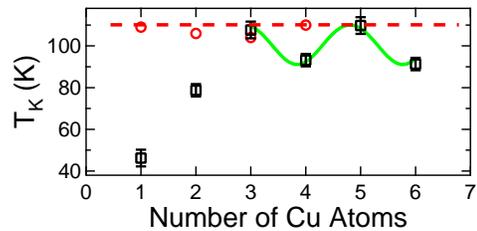}
  \caption{(Color online) $T_{\text{K}}$ of Co atoms in CoCu$_n$Co chains 
  as a function of $n$ (squares). A dashed line indicates 
  ${\tilde T}_{\text{K}}=110\,\text{K}$ of a single Co adatom attached to the 
  end of a Cu$_4$ chain. The circles (red) indicate calculated 
  \textit{single}-impurity Kondo temperatures, which would be expected for 
  CoCu$_n$Co chains in the absence of magnetic Co-Co interactions. The 
  sinusoidal line is a fit to data in the RKKY interaction regime with 
  calculated periodicity.}
  \label{fig2}
\end{figure}

\begin{figure*}
\begin{minipage}{.7\linewidth}
\centering
\includegraphics[width=.9\linewidth]{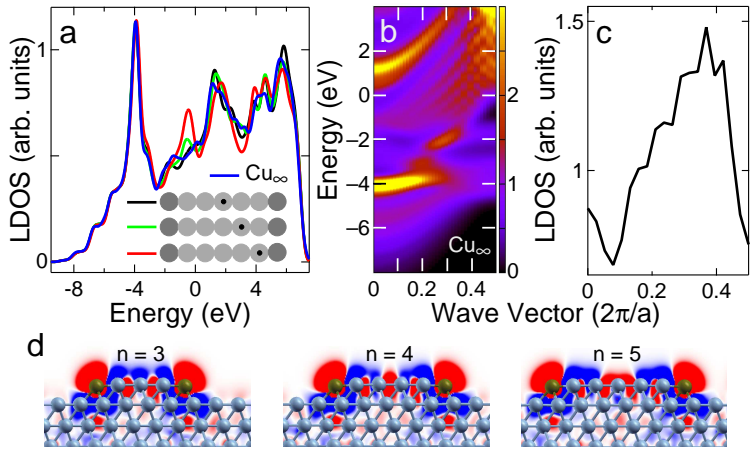}
\end{minipage}\begin{minipage}{.29\linewidth}
  \caption{(Color online) (a) Local density of states (LDOS) of Cu $4s$ and 
  $4p$ orbitals in a CoCu$_5$Co chain and in an infinitely long Cu chain 
  (Cu$_\infty$) on Cu(111). Cu sites at which the LDOS was evaluated are 
  indicated by dots. (b) LDOS of Cu$_\infty$ on Cu(111) versus energy and wave 
  vector along the chain direction. (c) Cross section of (b) at 
  $E=0\,\text{eV}$. (d) Relaxed CoCu$_n$Co ($3\leq n\leq 5$) chains with 
  calculated magnetization density for ferromagnetically aligned Co spins. 
  Light (red) and dark (blue) colors indicate up and down orientations, 
  respectively.}
  \label{fig3}
\end{minipage}
\end{figure*}

Linear clusters of Co and Cu atoms \cite{jla_07,jst_06,nne_08} were fabricated 
by manipulation with the STM tip by first assembling Cu$_{n}$ chains along the 
close-packed $[1\bar{1}0]$ direction and then 
attaching Co atoms (Fig.\,\ref{fig1}, left column). For $n \geq 3$, Co atoms 
are readily discriminated from Cu atoms by their larger apparent height. The 
right column of Fig.\,\ref{fig1} shows $\text{d}I/\text{d}V$ spectra (dots) 
acquired above the Co atoms of each CoCu$_n$Co chain. Spectra acquired at the 
two Co atoms were virtually identical.
The Abrikosov-Suhl resonance appears as an indentation of the $\text{d}I/\text{d}V$ signal around zero voltage and appreciably broadens from CoCuCo to CoCu$_3$Co.
To quantify its width, the spectra were fit by a Fano line \cite{ufa_61}, 
$\text{d}I/\text{d}V \propto (q+\epsilon)^2/(1+\epsilon^2)$ (black line). 
Fit parameters were the asymmetry parameter $q$ and  
$\epsilon =(\text{e}V-\epsilon_{\text{K}})/(\text{k}_{\text{B}} T_{\text{K}})$
($V$: sample voltage, $\epsilon_{\text{K}}$: resonance energy).
While $q\approx 0.1$ and $\epsilon_{\text{K}}\approx 2\,\text{meV}$ are rather independent of the number 
of Cu atoms, $T_{\text{K}}$ exhibits a pronounced variation 
(Fig.\,\ref{fig2}). $T_{\text{K}}$ nearly doubles from $n=1$ 
($T_{\text{K}}\approx 46\,\text{K}$) to $n=2$ ($\approx 79\,\text{K}$),   
increases further to $\approx 108\,\text{K}$ for CoCu$_3$Co, and then 
oscillates ($n=4$: $\approx 93\,\text{K}$, $n=5$: $\approx 110\,\text{K}$, 
$n=6$: $\approx 91\,\text{K}$). The maxima of the oscillation match the Kondo 
temperature of a Co atom at the end of CoCu$_3$ and CoCu$_4$ chains ($\tilde{T}_{\text{K}}=110\,\text{K}$), which approximate a 
CoCu$_\infty$ chain (dashed line in Fig.\,\ref{fig2}).

To interpret the evolution of $T_{\text{K}}$ a first-principles 
description of the CoCu$_n$Co chains on Cu(111) has been developed. DFT 
calculations were performed using the Vienna Ab Initio Simulation 
Package \cite{gkr_94} with the implemented Projector Augmented Waves basis 
sets \cite{pbl_94,gkr_99} and employing a generalized gradient approximation 
(GGA) \cite{Perdew:PW91} to the exchange correlation potential. The surface
was modeled by $(3\times 9)$ Cu(111) supercells with slab thicknesses of three to six layers and the CoCu$_n$Co chains oriented along the
$[1\bar{1}0]$ direction. The Kondo physics observed in the experiments are
governed by Co $3d$ electrons which are highly sensitive to their atomic 
environment \cite{nne_08}. Therefore, both  adatoms and first surface layer 
atoms were relaxed until the forces fell below 
$0.02\,\text{eV}\,\text{\AA}^{-1}$ for each atom. 

First, we show that \textit{single}-impurity physics cannot explain the experimental variation of $T_{\text{K}}$.
Using the Co hybridization functions obtained from DFT in a RG approach we estimated the variation of $T_{\text{K}}$ with cluster size due to single-impurity effects like variations in the local density of states at the Co site \cite{EPAPS}.
Extrapolating from $\tilde T_{\text{K}} = 110\,\text{K}$ for CoCu$_3$ and CoCu$_4$ to the CoCu$_n$Co chains yields variations 
of $T_{\text{K}}$ (Fig.\,\ref{fig2}, circles) which are much smaller than experimentally observed (Fig.\,\ref{fig2}, squares) and do not even follow the trend of the experimental data.

Due to the reduced coordination Cu atoms within the chain move by $0.16$ to 
$0.23\,\text{\AA}$ towards the Cu(111) surface, which corresponds to $8$ to 
$11\,\%$ of the (111) interlayer spacing. Co atoms at the ends of the chain 
move by $0.31$ to $0.33\,\text{\AA}$ toward Cu(111). These relaxations have a 
significant influence on the magnetic coupling between the Co atoms. This 
becomes clear from calculations of total energies of a CoCuCo chain for 
ferromagnetic ($\uparrow\uparrow$) and antiferromagnetic ($\uparrow\downarrow$)
spin orientations with and without geometric relaxation. The resulting exchange 
interaction energy, 
$E_{\text{ex}}=E_{\uparrow\downarrow}-E_{\uparrow\uparrow}$, is 
$14\pm 6\,\text{meV}$ for relaxed CoCuCo chains, while it is significantly 
smaller $|E_{\text{ex}}|\leq 1\,\text{meV}$ without relaxations. From a 
calculation within the local density approximation (LDA), 
$E_{\text{ex}}\approx 40\,\text{meV}$ was reported for the unrelaxed chain 
\cite{obr_08}. For relaxed CoCu$_2$Co chains we find 
$E_{\text{ex}}=-17\pm 3\,\text{meV}$, which is close to an earlier LDA result 
\cite{obr_08} and our GGA result $E_{\text{ex}}\approx -20\,\text{meV}$ for the 
unrelaxed chain. The signs of $E_{\text{ex}}$ indicate ferro- and 
antiferromagnetic coupling between the Co atoms in CoCuCo and CoCu$_2$Co 
chains, respectively.

Despite the ferromagnetic Co-Co coupling in the CoCuCo chain an Abrikosov-Suhl 
resonance with $T_{\text{K}}= 46\,\text{K}$ is observed in the 
experiments, which may appear surprising. The interplay of interimpurity 
coupling, $H=-J\vec{S}_1\cdot\vec{S}_2$, and Kondo screening has been 
theoretically addressed in terms of two-site spin-$1/2$ Kondo models 
\cite{cja_81,bjo_87,*bjo_88}. For ferromagnetic interimpurity exchange 
interaction, $J>0$, dominating over the single-impurity Kondo energy scale, 
$\text{k}_{\text{B}}\tilde{T}_{\text{K}}\ll J$, locking of the two spins to a 
total spin, $S_1+S_2=1$, followed by a spin-$1$ Kondo effect has been 
predicted \cite{cja_81}. In this case, the Kondo temperature is reduced, 
$T_{\text{K}}\approx\text{k}_{\text{B}}\tilde{T}_{\text{K}}^2/J$ \cite{foot01}. 
With $T_{\text{K}}=46\,\text{K}$ and $\tilde{T}_{\text{K}}=110\,\text{K}$, $J$ 
is estimated as $23\,\text{meV}$. As $J$ is related to 
$E_{\text{ex}}=14\pm 6\,\text{meV}$ by a factor of the order of one, the 
experimentally observed reduction of $\tilde{T}_{\text{K}}$ is well in line 
with $E_{\text{ex}}$ as obtained from our calculations \cite{foot02}. Consequently, theory and 
experiment consistently suggest that CoCuCo is in the crossover region between 
two independent and two ferromagnetically locked Kondo impurities 
\cite{cja_81,bjo_87,*bjo_88}, where a narrowed rather than a completely 
suppressed Abrikosov-Suhl resonance is found.

In the case of CoCu$_2$Co, where the Co atoms couple antiferromagnetically, our GGA 
calculations show that 
$|E_{\text{ex}}|\approx 2\,\text{k}_{\text{B}}\tilde{T}_{\text{K}}$. In a 
particle-hole-symmetric case, the two-site spin-$1/2$ Kondo model exhibits a 
quantum critical point at 
$J\approx 2.2\,\text{k}_{\text{B}}\tilde{T}_{\text{K}}$ separating ground 
states with an antiferromagnetically locked interimpurity singlet from two 
Kondo-screened impurities \cite{bjo_87,*bjo_88}. In systems without 
particle-hole symmetry, the quantum critical point is replaced by a crossover 
region, where the spectral weight of the Abrikosov-Suhl resonance is 
continuously reduced and evolves into a pseudogap feature 
\cite{osa_90,*osa_92a,*osa_92b,osa_93,iaf_95,jzh_10}. In this crossover region, there are two energy scales, $T_L<T_H$, 
characterizing the spin-fluctuations and the quasiparticle excitation spectra of the system \cite{osa_90,*osa_92a,*osa_92b,osa_93,jzh_10}.
The lower scale, $T_L$, 
gives rise to the sharpest and most pronounced feature in the spectral function 
at the Fermi level \cite{osa_93} and characterizes the onset of local Fermi 
liquid behavior \cite{jzh_10}. Hence, $\text{k}_{\text{B}}T_L$ should appear as 
the width of the experimentally observed Abrikosov-Suhl resonance and we refer 
to $T_L=T_{\text{K}}$ as the Kondo temperature of the two-impurity system. As 
$T_L<\tilde{T}_{\text{K}}$ \cite{osa_90,*osa_92a,*osa_92b,osa_93,jzh_10}, a 
narrowed Abrikosov-Suhl resonance as observed experimentally 
(Figs.\,\ref{fig1}, \ref{fig2}) is well in line with this crossover regime.
The results for $n=1,2$ prove that positive and negative exchange 
interactions lead to $T_{\text{K}}<\tilde{T}_{\text{K}}$ \cite{vir_89,vir_97}.

To understand the oscillations of $T_{\text{K}}$ for $n\geq 3$ the electronic 
structure of an infinite Cu chain (Cu$_\infty$) on Cu(111) is considered. Its 
structural relaxations (i.e. $0.2\,\text{\AA}$ downward relaxation of the chain atoms) and its local density of states (LDOS) are similar to 
those observed from CoCu$_n$Co for $n\geq 3$. Cu atoms that are no direct 
neighbors of Co atoms exhibit a similar LDOS as Cu atoms in Cu$_\infty$ 
[Fig.\,\ref{fig3}(a)]. Therefore, the Fermi wave vector $k_{\text{F}}$, which 
determines LDOS oscillations of the chains and the RKKY interaction, has been 
calculated from the band structure of Cu$_\infty$. Figure \ref{fig3}(b) shows 
the energy- and momentum-resolved LDOS of conduction electrons of 
Cu$_\infty$. The band of unoccupied states starting at $E=1.2$\,eV at $k=0$ has been 
measured in Refs.\,\cite{fol_04,jla_07} and does not contribute spectral 
weight at the Fermi energy ($E_{\text{F}}=0\,\text{eV}$). At $E_{\text{F}}$ 
[Fig.\,\ref{fig3}(c)] spectral weight is suppressed for 
$k_{\text{F}}<0.1\,(2\pi/a)$ [$a$: Cu(111) lattice constant, 
$a=2.57\,\text{\AA}$] and exhibits a maximum at $\approx 0.37\,(2\pi/a)$. 
Hence, Co--Co RKKY interactions and LDOS resonances at $E_{\text{F}}$ are 
expected to oscillate with a wave vector 
$2\,k_{\text{F}}\approx 0.74\,(2\pi/a)$ which by subtracting a reciprocal 
lattice vector is identical with $-0.26\,(2\pi/a)$ and corresponds to a direct 
space period of $\approx 3.8\,a$. This period can be clearly seen in 
Fig.\,\ref{fig3}(d) as an oscillatory magnetization density along the Cu 
chains. The period expected for $T_{\text{K}}$, however, is different. In the 
limit of weak RKKY interaction, the correction to the Kondo temperature reads 
$\tilde{T}_{\text{K}}^2-T_{\text{K}}^2\approx E_{\text{ex}}^2/\text{k}_{\text{B}}^2$
\cite{vir_97,vir_89,EPAPS}. Given that 
$E_{\text{ex}}^2\propto\sin^2(2\,k_{\text{F}}na)$ the spatial periodicity is 
reduced to $\approx 1.9\,a$, which corresponds well to the even-odd 
oscillations of $T_{\text{K}}$ observed in the experiments.

In summary, linking two Co atoms by a chain of Cu atoms nonlocal correlations 
between two Kondo impurities have been probed. The interimpurity interaction is proven to quench Kondo temperatures in short clusters and leads to RKKY 
induced oscillations at larger chain lengths. A reduction of the Kondo 
temperature independent of the sign of the interimpurity exchange interaction is found. These effects observed from a two impurity system may find a counterpart in crystalline solids as indicated by model studies of the double Bethe lattice \cite{Hafermann09}. 

Financial support by the Deutsche Forschungsgemeinschaft through SFB 668, the Innovationsfonds Schleswig-Holstein and FOM (The Netherlands) is acknowledged. 
We thank A. Rosch for providing his RG code.


\begin{thebibliography}{37}%
\makeatletter
\providecommand \@ifxundefined [1]{%
 \@ifx{#1\undefined}
}%
\providecommand \@ifnum [1]{%
 \ifnum #1\expandafter \@firstoftwo
 \else \expandafter \@secondoftwo
 \fi
}%
\providecommand \@ifx [1]{%
 \ifx #1\expandafter \@firstoftwo
 \else \expandafter \@secondoftwo
 \fi
}%
\providecommand \natexlab [1]{#1}%
\providecommand \enquote  [1]{``#1''}%
\providecommand \bibnamefont  [1]{#1}%
\providecommand \bibfnamefont [1]{#1}%
\providecommand \citenamefont [1]{#1}%
\providecommand \href@noop [0]{\@secondoftwo}%
\providecommand \href [0]{\begingroup \@sanitize@url \@href}%
\providecommand \@href[1]{\@@startlink{#1}\@@href}%
\providecommand \@@href[1]{\endgroup#1\@@endlink}%
\providecommand \@sanitize@url [0]{\catcode `\\12\catcode `\$12\catcode
  `\&12\catcode `\#12\catcode `\^12\catcode `\_12\catcode `\%12\relax}%
\providecommand \@@startlink[1]{}%
\providecommand \@@endlink[0]{}%
\providecommand \url  [0]{\begingroup\@sanitize@url \@url }%
\providecommand \@url [1]{\endgroup\@href {#1}{\urlprefix }}%
\providecommand \urlprefix  [0]{URL }%
\providecommand \Eprint [0]{\href }%
\providecommand \doibase [0]{http://dx.doi.org/}%
\providecommand \selectlanguage [0]{\@gobble}%
\providecommand \bibinfo  [0]{\@secondoftwo}%
\providecommand \bibfield  [0]{\@secondoftwo}%
\providecommand \translation [1]{[#1]}%
\providecommand \BibitemOpen [0]{}%
\providecommand \bibitemStop [0]{}%
\providecommand \bibitemNoStop [0]{.\EOS\space}%
\providecommand \EOS [0]{\spacefactor3000\relax}%
\providecommand \BibitemShut  [1]{\csname bibitem#1\endcsname}%
\let\auto@bib@innerbib\@empty
%</preamble>
\bibitem [{\citenamefont {Ruderman}\ and\ \citenamefont
  {Kittel}(1954)}]{mru_54}%
  \BibitemOpen
  \bibfield  {author} {\bibinfo {author} {\bibfnamefont {M.~A.}\ \bibnamefont
  {Ruderman}}\ and\ \bibinfo {author} {\bibfnamefont {C.}~\bibnamefont
  {Kittel}},\ }\href@noop {} {\bibfield  {journal} {\bibinfo  {journal} {Phys.
  Rev.}\ }\textbf {\bibinfo {volume} {96}},\ \bibinfo {pages} {99} (\bibinfo
  {year} {1954})}\BibitemShut {NoStop}%
\bibitem [{\citenamefont {Kasuya}(1956)}]{tka_56}%
  \BibitemOpen
  \bibfield  {author} {\bibinfo {author} {\bibfnamefont {T.}~\bibnamefont
  {Kasuya}},\ }\href@noop {} {\bibfield  {journal} {\bibinfo  {journal} {Prog.
  Theor. Phys.}\ }\textbf {\bibinfo {volume} {16}},\ \bibinfo {pages} {45}
  (\bibinfo {year} {1956})}\BibitemShut {NoStop}%
\bibitem [{\citenamefont {Yosida}(1957)}]{kyo_57}%
  \BibitemOpen
  \bibfield  {author} {\bibinfo {author} {\bibfnamefont {K.}~\bibnamefont
  {Yosida}},\ }\href@noop {} {\bibfield  {journal} {\bibinfo  {journal} {Phys.
  Rev.}\ }\textbf {\bibinfo {volume} {106}},\ \bibinfo {pages} {893} (\bibinfo
  {year} {1957})}\BibitemShut {NoStop}%
\bibitem [{\citenamefont {Jayaprakash}\ \emph {et~al.}(1981)\citenamefont
  {Jayaprakash}, \citenamefont {Krishnamurthy},\ and\ \citenamefont
  {Wilkins}}]{cja_81}%
  \BibitemOpen
  \bibfield  {author} {\bibinfo {author} {\bibfnamefont {C.}~\bibnamefont
  {Jayaprakash}}, \bibinfo {author} {\bibfnamefont {H.~R.}\ \bibnamefont
  {Krishnamurthy}}, \ and\ \bibinfo {author} {\bibfnamefont {J.~W.}\
  \bibnamefont {Wilkins}},\ }\href@noop {} {\bibfield  {journal} {\bibinfo
  {journal} {Phys. Rev. Lett.}\ }\textbf {\bibinfo {volume} {47}},\ \bibinfo
  {pages} {737} (\bibinfo {year} {1981})}\BibitemShut {NoStop}%
\bibitem [{\citenamefont {Jones}\ and\ \citenamefont {Varma}(1987)}]{bjo_87}%
  \BibitemOpen
  \bibfield  {author} {\bibinfo {author} {\bibfnamefont {B.~A.}\ \bibnamefont
  {Jones}}\ and\ \bibinfo {author} {\bibfnamefont {C.~M.}\ \bibnamefont
  {Varma}},\ }\href@noop {} {\bibfield  {journal} {\bibinfo  {journal} {Phys.
  Rev. Lett.}\ }\textbf {\bibinfo {volume} {58}},\ \bibinfo {pages} {843}
  (\bibinfo {year} {1987})}\BibitemShut {NoStop}%
\bibitem [{\citenamefont {Jones}\ \emph {et~al.}(1988)\citenamefont {Jones},
  \citenamefont {Varma},\ and\ \citenamefont {Wilkins}}]{bjo_88}%
  \BibitemOpen
  \bibfield  {author} {\bibinfo {author} {\bibfnamefont {B.~A.}\ \bibnamefont
  {Jones}}, \bibinfo {author} {\bibfnamefont {C.~M.}\ \bibnamefont {Varma}}, \
  and\ \bibinfo {author} {\bibfnamefont {J.~W.}\ \bibnamefont {Wilkins}},\
  }\href@noop {} {\bibfield  {journal} {\bibinfo  {journal} {Phys. Rev. Lett.}\
  }\textbf {\bibinfo {volume} {61}},\ \bibinfo {pages} {15} (\bibinfo {year}
  {1988})}\BibitemShut {NoStop}%
\bibitem [{\citenamefont {Zhou}\ \emph {et~al.}(2010)\citenamefont {Zhou},
  \citenamefont {Wiebe}, \citenamefont {Lounis}, \citenamefont {Vedmedenko},
  \citenamefont {Meier}, \citenamefont {Bl{\"{u}}gel}, \citenamefont
  {Dederichs},\ and\ \citenamefont {Wiesendanger}}]{lzh_10}%
  \BibitemOpen
  \bibfield  {author} {\bibinfo {author} {\bibfnamefont {L.}~\bibnamefont
  {Zhou}}, \bibinfo {author} {\bibfnamefont {J.}~\bibnamefont {Wiebe}},
  \bibinfo {author} {\bibfnamefont {S.}~\bibnamefont {Lounis}}, \bibinfo
  {author} {\bibfnamefont {E.}~\bibnamefont {Vedmedenko}}, \bibinfo {author}
  {\bibfnamefont {F.}~\bibnamefont {Meier}}, \bibinfo {author} {\bibfnamefont
  {S.}~\bibnamefont {Bl{\"{u}}gel}}, \bibinfo {author} {\bibfnamefont {P.~H.}\
  \bibnamefont {Dederichs}}, \ and\ \bibinfo {author} {\bibfnamefont
  {R.}~\bibnamefont {Wiesendanger}},\ }\href@noop {} {\bibfield  {journal}
  {\bibinfo  {journal} {Nature Phys.}\ }\textbf {\bibinfo {volume} {6}},\
  \bibinfo {pages} {187} (\bibinfo {year} {2010})}\BibitemShut {NoStop}%
\bibitem [{\citenamefont {Otte}\ \emph {et~al.}(2009)\citenamefont {Otte},
  \citenamefont {Ternes}, \citenamefont {Loth}, \citenamefont {Lutz},
  \citenamefont {Hirjibehedin},\ and\ \citenamefont {Heinrich}}]{ott_09}%
  \BibitemOpen
  \bibfield  {author} {\bibinfo {author} {\bibfnamefont {A.~F.}\ \bibnamefont
  {Otte}}, \bibinfo {author} {\bibfnamefont {M.}~\bibnamefont {Ternes}},
  \bibinfo {author} {\bibfnamefont {S.}~\bibnamefont {Loth}}, \bibinfo {author}
  {\bibfnamefont {C.~P.}\ \bibnamefont {Lutz}}, \bibinfo {author}
  {\bibfnamefont {C.~F.}\ \bibnamefont {Hirjibehedin}}, \ and\ \bibinfo
  {author} {\bibfnamefont {A.~J.}\ \bibnamefont {Heinrich}},\ }\href@noop {}
  {\bibfield  {journal} {\bibinfo  {journal} {Phys. Rev. Lett.}\ }\textbf
  {\bibinfo {volume} {103}},\ \bibinfo {pages} {107203} (\bibinfo {year}
  {2009})}\BibitemShut {NoStop}%
\bibitem [{\citenamefont {Uchihashi}\ \emph {et~al.}(2008)\citenamefont
  {Uchihashi}, \citenamefont {Zhang}, \citenamefont {Kr{\"{o}}ger},\ and\
  \citenamefont {Berndt}}]{tuc_08}%
  \BibitemOpen
  \bibfield  {author} {\bibinfo {author} {\bibfnamefont {T.}~\bibnamefont
  {Uchihashi}}, \bibinfo {author} {\bibfnamefont {J.}~\bibnamefont {Zhang}},
  \bibinfo {author} {\bibfnamefont {J.}~\bibnamefont {Kr{\"{o}}ger}}, \ and\
  \bibinfo {author} {\bibfnamefont {R.}~\bibnamefont {Berndt}},\ }\href@noop {}
  {\bibfield  {journal} {\bibinfo  {journal} {Phys. Rev. B}\ }\textbf {\bibinfo
  {volume} {78}},\ \bibinfo {pages} {033402} (\bibinfo {year}
  {2008})}\BibitemShut {NoStop}%
\bibitem [{\citenamefont {Bulla}\ \emph {et~al.}(2008)\citenamefont {Bulla},
  \citenamefont {Costi},\ and\ \citenamefont {Pruschke}}]{rbu_08}%
  \BibitemOpen
  \bibfield  {author} {\bibinfo {author} {\bibfnamefont {R.}~\bibnamefont
  {Bulla}}, \bibinfo {author} {\bibfnamefont {T.~A.}\ \bibnamefont {Costi}}, \
  and\ \bibinfo {author} {\bibfnamefont {T.}~\bibnamefont {Pruschke}},\
  }\href@noop {} {\bibfield  {journal} {\bibinfo  {journal} {Rev. Mod. Phys.}\
  }\textbf {\bibinfo {volume} {80}},\ \bibinfo {pages} {395} (\bibinfo {year}
  {2008})}\BibitemShut {NoStop}%
\bibitem [{\citenamefont {Chen}\ \emph {et~al.}(1999)\citenamefont {Chen},
  \citenamefont {Jamneala}, \citenamefont {Madhavan},\ and\ \citenamefont
  {Crommie}}]{wch_99}%
  \BibitemOpen
  \bibfield  {author} {\bibinfo {author} {\bibfnamefont {W.}~\bibnamefont
  {Chen}}, \bibinfo {author} {\bibfnamefont {T.}~\bibnamefont {Jamneala}},
  \bibinfo {author} {\bibfnamefont {V.}~\bibnamefont {Madhavan}}, \ and\
  \bibinfo {author} {\bibfnamefont {M.~F.}\ \bibnamefont {Crommie}},\
  }\href@noop {} {\bibfield  {journal} {\bibinfo  {journal} {Phys. Rev. B}\
  }\textbf {\bibinfo {volume} {60}},\ \bibinfo {pages} {R8529} (\bibinfo {year}
  {1999})}\BibitemShut {NoStop}%
\bibitem [{\citenamefont {Madhavan}\ \emph {et~al.}(2002)\citenamefont
  {Madhavan}, \citenamefont {Jamneala}, \citenamefont {Nagaoka}, \citenamefont
  {Chen}, \citenamefont {Li}, \citenamefont {Louie},\ and\ \citenamefont
  {Crommie}}]{vma_02}%
  \BibitemOpen
  \bibfield  {author} {\bibinfo {author} {\bibfnamefont {V.}~\bibnamefont
  {Madhavan}}, \bibinfo {author} {\bibfnamefont {T.}~\bibnamefont {Jamneala}},
  \bibinfo {author} {\bibfnamefont {K.}~\bibnamefont {Nagaoka}}, \bibinfo
  {author} {\bibfnamefont {W.}~\bibnamefont {Chen}}, \bibinfo {author}
  {\bibfnamefont {J.-L.}\ \bibnamefont {Li}}, \bibinfo {author} {\bibfnamefont
  {S.~G.}\ \bibnamefont {Louie}}, \ and\ \bibinfo {author} {\bibfnamefont
  {M.~F.}\ \bibnamefont {Crommie}},\ }\href@noop {} {\bibfield  {journal}
  {\bibinfo  {journal} {Phys. Rev. B}\ }\textbf {\bibinfo {volume} {66}},\
  \bibinfo {pages} {212411} (\bibinfo {year} {2002})}\BibitemShut {NoStop}%
\bibitem [{\citenamefont {Wahl}\ \emph {et~al.}(2007)\citenamefont {Wahl},
  \citenamefont {Simon}, \citenamefont {Diekh{\"{o}}ner}, \citenamefont
  {Stepanyuk}, \citenamefont {Bruno}, \citenamefont {Schneider},\ and\
  \citenamefont {Kern}}]{pwa_07}%
  \BibitemOpen
  \bibfield  {author} {\bibinfo {author} {\bibfnamefont {P.}~\bibnamefont
  {Wahl}}, \bibinfo {author} {\bibfnamefont {P.}~\bibnamefont {Simon}},
  \bibinfo {author} {\bibfnamefont {L.}~\bibnamefont {Diekh{\"{o}}ner}},
  \bibinfo {author} {\bibfnamefont {V.~S.}\ \bibnamefont {Stepanyuk}}, \bibinfo
  {author} {\bibfnamefont {P.}~\bibnamefont {Bruno}}, \bibinfo {author}
  {\bibfnamefont {M.~A.}\ \bibnamefont {Schneider}}, \ and\ \bibinfo {author}
  {\bibfnamefont {K.}~\bibnamefont {Kern}},\ }\href@noop {} {\bibfield
  {journal} {\bibinfo  {journal} {Phys. Rev. Lett.}\ }\textbf {\bibinfo
  {volume} {98}},\ \bibinfo {pages} {056601} (\bibinfo {year}
  {2007})}\BibitemShut {NoStop}%
\bibitem [{\citenamefont {Minamitani}\ \emph {et~al.}(2009)\citenamefont
  {Minamitani}, \citenamefont {Nakanishi}, \citenamefont
  {Di$\tilde{\texttt{n}}$o},\ and\ \citenamefont {Kasai}}]{emi_09}%
  \BibitemOpen
  \bibfield  {author} {\bibinfo {author} {\bibfnamefont {E.}~\bibnamefont
  {Minamitani}}, \bibinfo {author} {\bibfnamefont {H.}~\bibnamefont
  {Nakanishi}}, \bibinfo {author} {\bibfnamefont {W.~A.}\ \bibnamefont
  {Di$\tilde{\texttt{n}}$o}}, \ and\ \bibinfo {author} {\bibfnamefont
  {H.}~\bibnamefont {Kasai}},\ }\href@noop {} {\bibfield  {journal} {\bibinfo
  {journal} {J. Phys. Soc. Jpn.}\ }\textbf {\bibinfo {volume} {78}},\ \bibinfo
  {pages} {084705} (\bibinfo {year} {2009})}\BibitemShut {NoStop}%
\bibitem [{\citenamefont {Brovko}\ \emph {et~al.}(2008)\citenamefont {Brovko},
  \citenamefont {Ignatiev}, \citenamefont {Stepanyuk},\ and\ \citenamefont
  {Bruno}}]{obr_08}%
  \BibitemOpen
  \bibfield  {author} {\bibinfo {author} {\bibfnamefont {O.~O.}\ \bibnamefont
  {Brovko}}, \bibinfo {author} {\bibfnamefont {P.~A.}\ \bibnamefont
  {Ignatiev}}, \bibinfo {author} {\bibfnamefont {V.~S.}\ \bibnamefont
  {Stepanyuk}}, \ and\ \bibinfo {author} {\bibfnamefont {P.}~\bibnamefont
  {Bruno}},\ }\href@noop {} {\bibfield  {journal} {\bibinfo  {journal} {Phys.
  Rev. Lett.}\ }\textbf {\bibinfo {volume} {101}},\ \bibinfo {pages} {036809}
  (\bibinfo {year} {2008})}\BibitemShut {NoStop}%
\bibitem [{\citenamefont {Limot}\ \emph {et~al.}(2005)\citenamefont {Limot},
  \citenamefont {Kr{\"{o}}ger}, \citenamefont {Berndt}, \citenamefont
  {Garcia-Lekue},\ and\ \citenamefont {Hofer}}]{lli_05}%
  \BibitemOpen
  \bibfield  {author} {\bibinfo {author} {\bibfnamefont {L.}~\bibnamefont
  {Limot}}, \bibinfo {author} {\bibfnamefont {J.}~\bibnamefont {Kr{\"{o}}ger}},
  \bibinfo {author} {\bibfnamefont {R.}~\bibnamefont {Berndt}}, \bibinfo
  {author} {\bibfnamefont {A.}~\bibnamefont {Garcia-Lekue}}, \ and\ \bibinfo
  {author} {\bibfnamefont {W.~A.}\ \bibnamefont {Hofer}},\ }\href@noop {}
  {\bibfield  {journal} {\bibinfo  {journal} {Phys. Rev. Lett.}\ }\textbf
  {\bibinfo {volume} {94}},\ \bibinfo {pages} {126102} (\bibinfo {year}
  {2005})}\BibitemShut {NoStop}%
\bibitem [{\citenamefont {Lagoute}\ \emph {et~al.}(2007)\citenamefont
  {Lagoute}, \citenamefont {Nacci},\ and\ \citenamefont
  {F{\"{o}}lsch}}]{jla_07}%
  \BibitemOpen
  \bibfield  {author} {\bibinfo {author} {\bibfnamefont {J.}~\bibnamefont
  {Lagoute}}, \bibinfo {author} {\bibfnamefont {C.}~\bibnamefont {Nacci}}, \
  and\ \bibinfo {author} {\bibfnamefont {S.}~\bibnamefont {F{\"{o}}lsch}},\
  }\href@noop {} {\bibfield  {journal} {\bibinfo  {journal} {Phys. Rev. Lett.}\
  }\textbf {\bibinfo {volume} {98}},\ \bibinfo {pages} {146804} (\bibinfo
  {year} {2007})}\BibitemShut {NoStop}%
\bibitem [{\citenamefont {Stroscio}\ \emph {et~al.}(2006)\citenamefont
  {Stroscio}, \citenamefont {Tavazza}, \citenamefont {Crain}, \citenamefont
  {Celotta},\ and\ \citenamefont {Chaka}}]{jst_06}%
  \BibitemOpen
  \bibfield  {author} {\bibinfo {author} {\bibfnamefont {J.~A.}\ \bibnamefont
  {Stroscio}}, \bibinfo {author} {\bibfnamefont {F.}~\bibnamefont {Tavazza}},
  \bibinfo {author} {\bibfnamefont {J.~N.}\ \bibnamefont {Crain}}, \bibinfo
  {author} {\bibfnamefont {R.~J.}\ \bibnamefont {Celotta}}, \ and\ \bibinfo
  {author} {\bibfnamefont {A.~M.}\ \bibnamefont {Chaka}},\ }\href@noop {}
  {\bibfield  {journal} {\bibinfo  {journal} {Science}\ }\textbf {\bibinfo
  {volume} {313}},\ \bibinfo {pages} {948} (\bibinfo {year}
  {2006})}\BibitemShut {NoStop}%
\bibitem [{\citenamefont {N{\'{e}}el}\ \emph {et~al.}(2008)\citenamefont
  {N{\'{e}}el}, \citenamefont {Kr{\"{o}}ger}, \citenamefont {Berndt},
  \citenamefont {Wehling}, \citenamefont {Lichtenstein},\ and\ \citenamefont
  {Katsnelson}}]{nne_08}%
  \BibitemOpen
  \bibfield  {author} {\bibinfo {author} {\bibfnamefont {N.}~\bibnamefont
  {N{\'{e}}el}}, \bibinfo {author} {\bibfnamefont {J.}~\bibnamefont
  {Kr{\"{o}}ger}}, \bibinfo {author} {\bibfnamefont {R.}~\bibnamefont
  {Berndt}}, \bibinfo {author} {\bibfnamefont {T.~O.}\ \bibnamefont {Wehling}},
  \bibinfo {author} {\bibfnamefont {A.~J.}\ \bibnamefont {Lichtenstein}}, \
  and\ \bibinfo {author} {\bibfnamefont {M.~I.}\ \bibnamefont {Katsnelson}},\
  }\href@noop {} {\bibfield  {journal} {\bibinfo  {journal} {Phys. Rev. Lett.}\
  }\textbf {\bibinfo {volume} {101}},\ \bibinfo {pages} {266803} (\bibinfo
  {year} {2008})}\BibitemShut {NoStop}%
\bibitem [{\citenamefont {Fano}(1961)}]{ufa_61}%
  \BibitemOpen
  \bibfield  {author} {\bibinfo {author} {\bibfnamefont {U.}~\bibnamefont
  {Fano}},\ }\href@noop {} {\bibfield  {journal} {\bibinfo  {journal} {Phys.
  Rev.}\ }\textbf {\bibinfo {volume} {124}},\ \bibinfo {pages} {1866} (\bibinfo
  {year} {1961})}\BibitemShut {NoStop}%
\bibitem [{\citenamefont {Kresse}\ and\ \citenamefont {Hafner}(1994)}]{gkr_94}%
  \BibitemOpen
  \bibfield  {author} {\bibinfo {author} {\bibfnamefont {G.}~\bibnamefont
  {Kresse}}\ and\ \bibinfo {author} {\bibfnamefont {J.}~\bibnamefont
  {Hafner}},\ }\href@noop {} {\bibfield  {journal} {\bibinfo  {journal} {J.
  Phys.: Condens. Matter}\ }\textbf {\bibinfo {volume} {6}},\ \bibinfo {pages}
  {8245} (\bibinfo {year} {1994})}\BibitemShut {NoStop}%
\bibitem [{\citenamefont {Bl{\"{o}}chl}(1994)}]{pbl_94}%
  \BibitemOpen
  \bibfield  {author} {\bibinfo {author} {\bibfnamefont {P.~E.}\ \bibnamefont
  {Bl{\"{o}}chl}},\ }\href@noop {} {\bibfield  {journal} {\bibinfo  {journal}
  {Phys. Rev. B}\ }\textbf {\bibinfo {volume} {50}},\ \bibinfo {pages} {17953}
  (\bibinfo {year} {1994})}\BibitemShut {NoStop}%
\bibitem [{\citenamefont {Kresse}\ and\ \citenamefont
  {Joubert}(1999)}]{gkr_99}%
  \BibitemOpen
  \bibfield  {author} {\bibinfo {author} {\bibfnamefont {G.}~\bibnamefont
  {Kresse}}\ and\ \bibinfo {author} {\bibfnamefont {D.}~\bibnamefont
  {Joubert}},\ }\href@noop {} {\bibfield  {journal} {\bibinfo  {journal} {Phys.
  Rev. B}\ }\textbf {\bibinfo {volume} {59}},\ \bibinfo {pages} {1758}
  (\bibinfo {year} {1999})}\BibitemShut {NoStop}%
\bibitem [{\citenamefont {Perdew}\ \emph {et~al.}(1992)\citenamefont {Perdew}
  \emph {et~al.}}]{Perdew:PW91}%
  \BibitemOpen
  \bibfield  {author} {\bibinfo {author} {\bibfnamefont {J.~P.}\ \bibnamefont
  {Perdew}} \emph {et~al.},\ }\href {\doibase 10.1103/PhysRevB.46.6671}
  {\bibfield  {journal} {\bibinfo  {journal} {Phys. Rev. B}\ }\textbf {\bibinfo
  {volume} {46}},\ \bibinfo {pages} {6671} (\bibinfo {year}
  {1992})}\BibitemShut {NoStop}%
\bibitem [{EPA()}]{EPAPS}%
  \BibitemOpen
  \href@noop {} {}\bibinfo {note} {See EPAPS Document No. XXX for details of
  the perturbative RG analysis and the generalization of \cite{vir_89} to the
  two-impurity Kondo problem.}\BibitemShut {Stop}%
\bibitem [{foo({\natexlab{a}})}]{foot01}%
  \BibitemOpen
  \href@noop {} {}\bibinfo {note} {This is a limiting
  case of Footnote 10 in \cite{cja_81} for $k_{\text{F}}R\gg 1$ or
  $k_{\text{F}}R\approx n\pi$ with integer nonzero $n$.}\BibitemShut {Stop}%
\bibitem [{foo({\natexlab{b}})}]{foot02}%
  \BibitemOpen
  \href@noop {} {}\bibinfo {note} {{T}his very good
  agreement should be considered only qualitatively, as
  {$T_{\text{K}}\approx\text{k}_{\text{B}}\tilde{T}_{\text{K}}^2/J$} has been
  derived for {$J\gg\text{k}_{\text{B}}\tilde{T}_{\text{K}}$}, while here we
  have {$J\approx 2\,\text{k}_{\text{B}}{\tilde T}_{\text{K}}$}.}\BibitemShut
  {Stop}%
\bibitem [{\citenamefont {Sakai}\ \emph {et~al.}(1990)\citenamefont {Sakai},
  \citenamefont {Shimizu},\ and\ \citenamefont {Kasuya}}]{osa_90}%
  \BibitemOpen
  \bibfield  {author} {\bibinfo {author} {\bibfnamefont {O.}~\bibnamefont
  {Sakai}}, \bibinfo {author} {\bibfnamefont {Y.}~\bibnamefont {Shimizu}}, \
  and\ \bibinfo {author} {\bibfnamefont {T.}~\bibnamefont {Kasuya}},\
  }\href@noop {} {\bibfield  {journal} {\bibinfo  {journal} {Sol. State
  Commun.}\ }\textbf {\bibinfo {volume} {75}},\ \bibinfo {pages} {81} (\bibinfo
  {year} {1990})}\BibitemShut {NoStop}%
\bibitem [{\citenamefont {Sakai}\ and\ \citenamefont
  {Shimizu}(1992{\natexlab{a}})}]{osa_92a}%
  \BibitemOpen
  \bibfield  {author} {\bibinfo {author} {\bibfnamefont {O.}~\bibnamefont
  {Sakai}}\ and\ \bibinfo {author} {\bibfnamefont {Y.}~\bibnamefont
  {Shimizu}},\ }\href@noop {} {\bibfield  {journal} {\bibinfo  {journal} {J.
  Phys. Soc. Jpn.}\ }\textbf {\bibinfo {volume} {61}},\ \bibinfo {pages} {2333}
  (\bibinfo {year} {1992}{\natexlab{a}})}\BibitemShut {NoStop}%
\bibitem [{\citenamefont {Sakai}\ and\ \citenamefont
  {Shimizu}(1992{\natexlab{b}})}]{osa_92b}%
  \BibitemOpen
  \bibfield  {author} {\bibinfo {author} {\bibfnamefont {O.}~\bibnamefont
  {Sakai}}\ and\ \bibinfo {author} {\bibfnamefont {Y.}~\bibnamefont
  {Shimizu}},\ }\href@noop {} {\bibfield  {journal} {\bibinfo  {journal} {J.
  Phys. Soc. Jpn.}\ }\textbf {\bibinfo {volume} {61}},\ \bibinfo {pages} {2348}
  (\bibinfo {year} {1992}{\natexlab{b}})}\BibitemShut {NoStop}%
\bibitem [{\citenamefont {Sakai}\ \emph {et~al.}(1993)\citenamefont {Sakai},
  \citenamefont {Shimizu},\ and\ \citenamefont {Kaneko}}]{osa_93}%
  \BibitemOpen
  \bibfield  {author} {\bibinfo {author} {\bibfnamefont {O.}~\bibnamefont
  {Sakai}}, \bibinfo {author} {\bibfnamefont {Y.}~\bibnamefont {Shimizu}}, \
  and\ \bibinfo {author} {\bibfnamefont {N.}~\bibnamefont {Kaneko}},\
  }\href@noop {} {\bibfield  {journal} {\bibinfo  {journal} {Physica B}\
  }\textbf {\bibinfo {volume} {186}},\ \bibinfo {pages} {323} (\bibinfo {year}
  {1993})}\BibitemShut {NoStop}%
\bibitem [{\citenamefont {Affleck}\ \emph {et~al.}(1995)\citenamefont
  {Affleck}, \citenamefont {Ludwig},\ and\ \citenamefont {Jones}}]{iaf_95}%
  \BibitemOpen
  \bibfield  {author} {\bibinfo {author} {\bibfnamefont {I.}~\bibnamefont
  {Affleck}}, \bibinfo {author} {\bibfnamefont {A.~W.~W.}\ \bibnamefont
  {Ludwig}}, \ and\ \bibinfo {author} {\bibfnamefont {B.~A.}\ \bibnamefont
  {Jones}},\ }\href@noop {} {\bibfield  {journal} {\bibinfo  {journal} {Phys.
  Rev. B}\ }\textbf {\bibinfo {volume} {52}},\ \bibinfo {pages} {9528}
  (\bibinfo {year} {1995})}\BibitemShut {NoStop}%
\bibitem [{\citenamefont {Zhu}\ and\ \citenamefont {Zhu}(2010)}]{jzh_10}%
  \BibitemOpen
  \bibfield  {author} {\bibinfo {author} {\bibfnamefont {L.}~\bibnamefont
  {Zhu}}\ and\ \bibinfo {author} {\bibfnamefont {J.-X.}\ \bibnamefont {Zhu}},\
  }\href@noop {} {\bibfield  {journal} {\bibinfo  {journal} {arxiv:1005.5154}\
  } (\bibinfo {year} {2010})}\BibitemShut {NoStop}%
\bibitem [{\citenamefont {Irkhin}\ and\ \citenamefont
  {Katsnelson}(1989)}]{vir_89}%
  \BibitemOpen
  \bibfield  {author} {\bibinfo {author} {\bibfnamefont {V.~Y.}\ \bibnamefont
  {Irkhin}}\ and\ \bibinfo {author} {\bibfnamefont {M.~I.}\ \bibnamefont
  {Katsnelson}},\ }\href@noop {} {\bibfield  {journal} {\bibinfo  {journal} {Z.
  Phys. B}\ }\textbf {\bibinfo {volume} {75}},\ \bibinfo {pages} {67} (\bibinfo
  {year} {1989})}\BibitemShut {NoStop}%
\bibitem [{\citenamefont {Irkhin}\ and\ \citenamefont
  {Katsnelson}(1997)}]{vir_97}%
  \BibitemOpen
  \bibfield  {author} {\bibinfo {author} {\bibfnamefont {V.~Y.}\ \bibnamefont
  {Irkhin}}\ and\ \bibinfo {author} {\bibfnamefont {M.~I.}\ \bibnamefont
  {Katsnelson}},\ }\href@noop {} {\bibfield  {journal} {\bibinfo  {journal}
  {Phys. Rev. B}\ }\textbf {\bibinfo {volume} {56}},\ \bibinfo {pages} {8109}
  (\bibinfo {year} {1997})}\BibitemShut {NoStop}%
\bibitem [{\citenamefont {F\"olsch}\ \emph {et~al.}(2004)\citenamefont
  {F\"olsch}, \citenamefont {Hyldgaard}, \citenamefont {Koch},\ and\
  \citenamefont {Ploog}}]{fol_04}%
  \BibitemOpen
  \bibfield  {author} {\bibinfo {author} {\bibfnamefont {S.}~\bibnamefont
  {F\"olsch}}, \bibinfo {author} {\bibfnamefont {P.}~\bibnamefont {Hyldgaard}},
  \bibinfo {author} {\bibfnamefont {R.}~\bibnamefont {Koch}}, \ and\ \bibinfo
  {author} {\bibfnamefont {K.~H.}\ \bibnamefont {Ploog}},\ }\href {\doibase
  10.1103/PhysRevLett.92.056803} {\bibfield  {journal} {\bibinfo  {journal}
  {Phys. Rev. Lett.}\ }\textbf {\bibinfo {volume} {92}},\ \bibinfo {pages}
  {056803} (\bibinfo {year} {2004})}\BibitemShut {NoStop}%
\bibitem [{\citenamefont {Hafermann}\ \emph {et~al.}(2009)\citenamefont
  {Hafermann}, \citenamefont {Katsnelson},\ and\ \citenamefont
  {Lichtenstein}}]{Hafermann09}%
  \BibitemOpen
  \bibfield  {author} {\bibinfo {author} {\bibfnamefont {H.}~\bibnamefont
  {Hafermann}}, \bibinfo {author} {\bibfnamefont {M.~I.}\ \bibnamefont
  {Katsnelson}}, \ and\ \bibinfo {author} {\bibfnamefont {A.~I.}\ \bibnamefont
  {Lichtenstein}},\ }\href@noop {} {\bibfield  {journal} {\bibinfo  {journal}
  {Europhys. Lett.}\ }\textbf {\bibinfo {volume} {85}},\ \bibinfo {pages}
  {37006} (\bibinfo {year} {2009})}\BibitemShut {NoStop}%
\end{thebibliography}
\end{document}